\documentclass[12pt]{article}
\usepackage[dvipdfmx]{graphicx, color}
\usepackage{amsfonts}
\usepackage{amsmath}
\usepackage{amscd}
\usepackage{amssymb}
\usepackage{bm}
\usepackage{bbm}
\usepackage[all]{xy}
\usepackage{enumerate}

\newcommand{\ket}[1]{\left| #1\right\rangle}
\newcommand{\bracket}[3]{\left\langle #1\right|#2 \left| #3\right\rangle}

\begin{document} 

\begin{center}
{\Large \bf Casimir energy between two parallel plates\\
and projective representation of Poincar\'{e }group 
 }

\vspace{15mm}

Takamaru Akita $^{\rm{1}}$
            \footnote{E-mail: akita@q.phen.mie-u.ac.jp}
and \; 
Mamoru Matsunaga $^{\rm{1},  \rm{2}}$
            \footnote{E-mail: matsuna@phen.mie-u.ac.jp; matsunaga@ars.mie-u.ac.jp}
\end{center}

\begin{center}
\textit{
$^{\rm{1}}$Department of Physics Engineering, Mie University, \\
     Tsu 514-8507, JAPAN
    \\
$^{\rm{2}}$College of Liberal Arts and Sciences, 
     Mie University, \\
     Tsu 514-8507, JAPAN}
\end{center}

\vspace{5mm}

\begin{abstract}
The Casimir effect is a physical manifestation of  zero point energy of quantum vacuum.  In a relativistic quantum field theory, Poincar\'{e } symmetry of the theory seems,  at first sight, to imply that non-zero vacuum energy is inconsistent with translational invariance of the vacuum. 
In the setting of two uniform boundary plates at rest, quantum fields outside the plates have (1+2)-dimensional Poincar\'{e }symmetry.
Taking a massless scalar field as an example, we have examined the consistency between the Poincar\'{e }symmetry and the existence of the vacuum energy.
We note that, in quantum theory, symmetries are represented projectively in general and  show that the Casimir energy is connected to central charges appearing in the algebra of generators in the projective representations.
 \end{abstract}

\newpage 
\section{Introduction}
Since the pioneering work of Casimir \cite{Casimir},  vacuum energy of quantum fields  has been the subject of intense investigations from both experimental and theoretical sides~\cite{Plunien/Phys. Rep., Dalvit, Oxford, Milton}. Experimental measurements of the Casimir forces, by using an atomic force microscope  or micro-electromechanical system,  reach the high precision at the level within 1\% and agreement with the theoretical prediction is also at the same level at least for zero temperature.
Theoretical investigation of the Casimir effects extends a variety of fields of physics such as particle physics,  atomic physics, astrophysics and cosmology, and condensed matter physics.  In particle physics, for example, the Casimir energy of quark and gluon fields inside a hadron makes essential contributions to its mass. The Casimir force offers one of the effective mechanisms for spontaneous compactification of extra spatial dimensions in the Kaluza-Klein theories. 

This paper discuss more theoretical issue, i.e. we examine the consistency between the existence of the Casimir energy and the Poincar\'{e }symmetry in the setting of two uniform perfectly-reflecting parallel boundary planes at rest. In this configuration, the quantum field  theory is invariant under the time-translation, the translations and boosts along the plane, and under the rotation in the plane. As a result of these invariances of the theory, it seems that, if require the translational invariance of the vacuum (vanishing total momentum of the field), then the vacuum energy should  vanish. This argument has a loophole as expected. 
We pay our attention to the representation of symmetries in quantum theory and to the fact that, in order to compare the zero point energies, we have to consider time-dependent Hamiltonian connecting the different static configurations. 

The paper is organize as follows. In Sec. 2, we set up the problem in the example of a massless scalar field. Sec. 3 summarizes the projective representation and linear representations of symmetries in quantum theory. In Sec. 4, an adiabatic process connecting  two static configuration is analyzed. The final section is devoted to the conclusion.

\section{Setup}
In this section，we set up the problem in the case of massless scalar field theory in $O\mathchar`- x_1 x_2 x_3$ space with the Dirichlet boundary condition $\phi = 0$ on the planes $x_3 = - L , L$.

This theory is invariant under the translations along the $x_1\mathchar`-, x_2\mathchar`-$directions and the time direction,
under the rotation in $x_1 x_2\mathchar`-$plane, and under the boosts along the $x_1\mathchar`-, x_2\mathchar`-$directions. These transformations and their compositions form the (1+2)-dimensional Poincar\'{e } group. Denoting the respective generator as $P_1, P_2, H, J, K_1, K_2$, we are given the Lie algebra of the Poincar\'{e } group 
 \begin{align}
 [P_i, P_j] &= [P_i, H] = [J,  H] =  0, 	\label{Poi1} \\
 i [J, P_i] &= - \epsilon_{ij} P^j , \quad\quad	  i [J, K_i] =  \epsilon_{ij} K^j ,	\label{Poi2}\\
 i  [K_i, K_j] &=  \epsilon_{ij}  J,	\quad\quad\quad i [K_i, H ] =  P_i,	\label{Poi3}\\
  i [K_i, P_j ] &=  \delta_{ij} H, 	\label{Poi4}
\end{align}
with $i, j = 1, 2$,  the antisymmetric tensor $\epsilon_{12}=1$ and summation for the repeated indices being implied.

Since the field $\phi$ satisfies the d'Alembert equation and the Dirichlet boundary condition, $\phi$ can be expanded as
\[
\phi (t, x_1, x_2, x_3) = \sum_{n = 1}^\infty \phi^{(n)} (t, x_1, x_2) \sin \frac{n \pi }{L}x_3, 
\]
where  the expansion coefficient $\phi^{(n)} $ satisfies the (1+2)-dimensional Klein-Gordon equation
\[
(\square_2 +  {m_n}^2 )\phi^{(n)} (t, x_1, x_2)  = 0
\quad (n=1, 2, \ldots).
\]
with $m_n := n \pi /L$. i.e. each mode $\phi^{(n)} $ forms (1+2)-dimensional scalar field with the mass $m_n $.

The Lagrangian density $\mathcal{L}$ for the (1+3)-dimensional field $\phi$ is written as the sum of the Lagrangian density  $\mathcal{L}^{(n)}$ for the (1+2)-dimensional fields $\phi^{(n)}$:
\begin{gather}
\mathcal{L} =  \sum_{n = 1}^\infty \mathcal{L}^{(n)} , \label{Lag1}\\
\mathcal{L}^{(n)} =\frac{1}{2} (\partial^\mu \phi^{(n)}  \partial_\mu \phi^{(n)} - {m_n}^2  \phi^{(n)\,2}). \label{Lag2}
\end{gather}
As a result, the  (1+3)-dimensional Poincar\'{e } algebra is decomposed to a direct sum of the  (1+2)-dimensional ones generated by the generators $P^{(n)}_i ,  H^{(n)}, J^{(n)}, K^{(n)}_i$ for each $n$-th mode.
These generators have the same form of commutation relation as eqs.~\eqref{Poi1}--\eqref{Poi4}.
In the following we take up the $n$-th mode and drop the upperscript $(n)$.

We express the generators of the  Poincar\'{e } algebra for the $n$-th mode  in terms of the canonical conjugate pairs $\phi (t, \bm{x}), \pi (t, \bm{x})\;(\bm x:=(x_1, x_2) )$, whose dynamics is derive from the Lagrangian~\eqref{Lag2} and whose commutation relations are  given by
\begin{align}
[\phi (t, \bm{x}), \phi (t, \bm{x}^\prime)] &= [\pi(t, \bm{x}), \pi(t, \bm{x}^\prime)] = 0, 	\label{CCR1}\\
[\phi (t, \bm{x}), \pi (t, \bm{x}^\prime)] &= i \delta^2 (\bm{x} - \bm{x}^\prime) 	 	\label{CCR2}.
\end{align}

Following Noether's prescription, we obtain

\begin{align}
H &= \int d^2x \,\mathcal{H}, \quad\quad \mathcal{H} = \frac{1}{2}\left(\pi^2 + (\nabla \phi)^2 +  {m_n}^2 \phi^2 \right),\label{Hc}\\
P_i &= \int d^2x\, \pi\, \partial_i \phi,  \quad\quad K_i = t P_i - \int d^2x \,x_i\mathcal{H}, \label{Pc}\\
J &= \int d^2x\, \epsilon_{ij} x^i \,\pi\, \partial^j \phi.
\end{align}
Using the canonical commutation relations~\eqref{CCR1}, \eqref{CCR2}, we see that  eqs.~\eqref{Poi1}--\eqref{Poi4} are satisfied.
In particular, the vacuum expectation values of eq.~\eqref{Poi4}
\begin{equation}
i [K_i, P_i] = H \quad (i = 1, 2) \label{Poi4i}
\end{equation}
gives
\begin{equation}
i\bracket{0}{ [K_i, P_i]}{0} = \bracket{0}{H}{0} .  \label{H0}
\end{equation}
If we require the translational invariance of the vacuum, $P \ket{0} =0$, then eqs.~\eqref{Hc} and \eqref{H0} give
\begin{equation}
\bracket{0}{H}{0} =  \frac{1}{2}\int d^2x \left(\bracket{0}{\pi^2 }{0}+ \bracket{0}{(\nabla \phi)^2}{0} +  {m_n}^2 \bracket{0}{\phi^2 }{0}\right) =0,
\end{equation}
which means
\[
\phi \ket{0} = \nabla \phi \ket{0} = \pi \ket{0} = 0
\]
in contradiction to eq.~\eqref{CCR2}.

Usually, in field theories without boundaries, with the aid of  the arbitrary additive constant inherent in the definition of the Hamiltonian, $H$ is redefined to satisfy $H\ket{0}=0$, which, in turn, seems to mean the nonexistence of Casimir energy. This is nothing but the inconsistency sketched out in Sec. 1.

In the following sections, we show that this apparent inconsistency disappears if we note the following two points:
\begin{itemize}
\item In quantum theory, symmetries are represented \textit{projectively} in general, and represented linearly if certain condition is satisfied.
\item In the Casimir effect, comparison between the vacuum energies of two different configuration are made: in the setting we are considering, two configuration of the plates are  e.g. $L = L_0$ and $L=L_1$.  Since the system should be described by a single Hamiltonian, we are to consider time-dependent Hamiltonian connecting $L = L_0$ and $L=L_1$.
\end{itemize}

\section{Projective representation of Poincar\'{e }group}
In Sec. 2 we describe the Lie algebra of Poincar\'{e }group as eqs.~\eqref{Poi1}--\eqref{Poi4}.  If the group is linearly represented, i.e. represented by a homomorphism from the group to linear operators, then the Lie algebra is nothing but the commutator algebra of the generators.
 However, as is well-known, in quantum theory, symmetry group $G$ is represented projectively in general~\cite{Weinberg}:  Unitary operators $U(g)\; (g \in G)$ form a linear representation up to phase factor, which means 
\begin{equation}
U(g)\,U(g^\prime) = e^{i\theta (g, g^\prime)}U(g\, g^\prime) 	\quad (g, g^\prime \in G).   \label{PRREP}
\end{equation}
Setting  $U(e) = I$ without loss of generality and expanding $U(g)$ around $g = e$, we get, from eq.~\eqref{PRREP}, the algebra of the generators, wherein there appear central charges corresponding to the phase factor $e^{i\theta}$. The associative law of the products of $U(g)$'s
\begin{equation}
U(g)\,\left( U(g^\prime)\, U(g^{\prime \prime}) \right)= \left( U(g)\,U(g^\prime)\right)\, U(g^{\prime \prime})	\quad ( g, g^\prime, g^{\prime\prime} \in G)	\label{asso}
\end{equation}
gives some constraints, called the cocycle condition, on the the phases $\theta (g, g^\prime)$.
If we multiply $U(g)$ by a phase factor $e^{i \alpha (g)}$ and redefine $e^{i \alpha (g)}\, U(g)$ as  $U(g)$，then the phase becomes $\theta (g, g^\prime) - \alpha (g) -\alpha (g^\prime)$. In most cases, this redefinition of $U(g)$ could makes the phase factor to disappear~\cite{Azcarraga,Bargmann}.

In our case of (1+2)-dimensional  Poincar\'{e }group, the algebra of the generators $P_1, P_2, H, J, K_1, K_2$ has 
the central charges in the righthand side of eq.~\eqref{Poi1}--\eqref{Poi4}: for example,
\begin{align}
i [K_i, H ] &=  P_i + C_{0, 0i},		\label{projKH}\\
 i [K_i, P_j ] &=  \delta_{ij} H 	+ C_{j, 0i}	,	\label{projKP}\\
 i [J, P_i] &= - \epsilon_{ij} P^j + C_{i, 12}.	 \label{projJP}
\end{align}
The cocycle condition for the central charges is
\begin{gather}
C_{\mu, \lambda \nu} =  g_{\mu \nu}\,C_{\lambda} -  g_{\mu \lambda}\,C_{\nu}, 	\label{C1}\\
C_{\lambda}  :=  \frac{1}{2} g^{\mu \nu} C_{\mu, \lambda \nu}.	\label{C2}
\end{gather}
From these equations, we get
\begin{gather}
C_{0, 0i} = - C_{i}, 	\label{CC1}\\
C_{j, 0i} =  - \delta_{ij} C_0, 	\label{CC2}\\
C_{i, 12} =  - \epsilon_{ij} C^j	\label{CC3}.
\end{gather}
Thus, we can eliminate $C_i$ and $C_0$ by redefining $P_i + C_i$ and $H + C_0$ as $P_i $ and $H$, respectively. Other central charges disappear by similar redefinitions of   $J$ and $K_i$.

The choice of  arbitrary additive constant  in the definition of the Hamiltonian, mentioned in Sec. 2, corresponds to the  elimination of the central charge $C_0$. 

As for the setting of two parallel plates discussed in this paper, even if we eliminate the central charge in one configuration, there remains non-zero central charge in the other configuration.

\section{Adiabatic process and projective representation}
The Casimir energy is the difference between the vacuum energies of two different configurations, in our case $L = L_0$ and $L=L_1$. The Hamiltonian of the $n$-th mode scalar field given by
\begin{equation}
H(L) = \int d^2x \,\mathcal{H}(L) = \frac{1}{2}\int d^2x  \left(\pi^2 + (\nabla \phi)^2 +  {m^2_n(L)} \phi^2 \right) \label{H(L)}
\end{equation}
has the mass parameter $m_n (L) = n\pi/L$ and hence become time-dependent when connecting these two configurations.

We set the time-dependence of $L$ as, for example, 
\begin{gather}
s_T  (t) = \tanh(\tan \frac{\pi t}{2 T}),\\
L(t) =  \begin{cases}   L_0 & (t \leq - T)\\
	  			\displaystyle{\frac{1- s_T  (t) }{2} }L_0 +\displaystyle{ \frac{1+ s_T  (t) }{2} }L_1 & (  - T \leq t \leq  T) \\
  				L_1 &  (T \leq t)
	 	  \end{cases}
.
\end{gather}
as shown in Fig. 1.
\begin{figure}[h]
\begin{center}
\includegraphics[width=7.5cm] {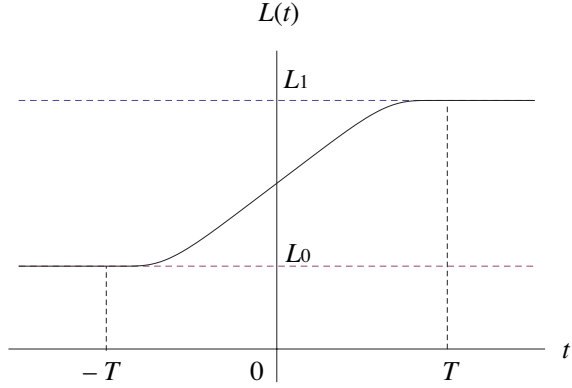}
\caption{Adiabatic change of the distance $L(t)$}
\label{Fig}
\end{center}
\end{figure}

In the regions  $t \leq - T$ and $ t  \geq  T$, the Hamiltonian, denoted as $H(L_0)$ and $H(L_1)$ respectively,  is  time-independent,  and hence  the system is invariant under the infinitesimal Poincar\'{e } transformations.

We denote the vacuum state of $H(L_0)$ as $\ket{0}_{L_0}$ and the energy eigenvalue as $E_0$:
\begin{equation}
H(L_0) \ket{0}_{L_0} = E_0 \ket{0}_{L_0}. \label{E0}
\end{equation}

In the region $- T \leq t \leq  T$, the state $\ket{0}_{L_0}$ is not necessarily an eigenstate of the Hamiltonian $H(L)$. However, because the excited states of $H(L)$ given by Eq.~\eqref{H(L)} consist of quanta with the mass $m_n(L) $, excitation energy above the ground state is greater than $m_n(L) > 0$.  Hence  we can  invoke the adiabatic theorem:
If we take $T$ large enough, then the state $\ket{0}_{L_0}$ remains the ground state of $H(L)$, which means in particular
\begin{equation}
H(L_1) \ket{0}_{L_0} = E_1 \ket{0}_{L_0} . \label{E1}
\end{equation}

Now, we consider, in the regions $t \leq - T$ and $t  \geq  T$,  the consistency of  invariance under the infinitesimal Poincar\'{e } transformations and the existence of the Casimir energy.
We take the generators for the translations and the boosts as 
\begin{gather}
P_i =  \int d^2x\, :\pi\, \partial_i \phi : \label{:P:}\\
K_i(L) = t P_i - \int d^2x \,x_i \mathcal{H}(L) \label{K(L)}.
\end{gather}
The generators $P_i$ are independent of $L$ and are defined by normal ordered product, while the generators $H(L)$ given by Eq.~\eqref{H(L)} and $K_i(L) $ are dependent on $L$ and hence are not normal-ordered.
Straightforward calculation of the left-hand side of eq.~\eqref{projKP} for $i=j$ shows that
\begin{equation}
i [K_{i}(L), P_i] = H(L) - E(L),  \label{KP}
\end{equation}
where $E(L)$ is an $L$-dependent constant.
From eqs.~\eqref{E0}, \eqref{E1}, \eqref{KP}, we see that, if we choose $E_0 = E(L_0)$ and $E_1 = E(L_1)$, then the translational invariance of the vacuum $\ket{0}_{L_0}$
\[
P_i\ket{0}_{L_0} = P_i\ket{0}_{L_1} = 0
\]
and the commutator algebra with the central charges of Poincar\'{e } generators
\begin{gather}
i [K_{i}(L_0), P_i] = H(L_0) - E_0  \label{K(L0)}\\
i [K_{i}(L_1), P_i] = H(L_1) - E_1 \label{K(L1)}
\end{gather}
are compatible, in contrast to eq.~\eqref{H0}.

By adding constant $E_0$ to $H(L)$,  we could redefine $H(L)$ so as to eliminate central charge in eq.~\eqref{K(L0)}, i.e. we could  reproduce eq.~\eqref{Poi4i} in the region $t \leq - T$.  Then,  in the region $t  \geq  T$, there remains central charge $E_0 -E_1$ in eq.~\eqref{K(L1)}.
\section{Summary}
In this paper, we have confirmed the consistency between the existence of the Casimir energy and translational invariance of the vacuum of the Poincar\'{e } invariant massless scalar field in the configuration of two parallel boundary plates. The points are:
\begin{itemize}
\item  Since, in the Casimir effect, comparison between the vacuum energies of two different static configuration are made,  we are to consider time-dependent Hamiltonian connecting these static configurations.

\item Even if we could choose the additive constant of the Hamiltonian so as to make the representation of the Poincar\'{e } group linear (no central charge in the algebra) in the one configuration, the representation of the group become projective (nonzero central charge in the algebra) in the other.
\end {itemize}
A few comments are order.
First, the additive constants such as $E(L)$ in Eq.~\eqref{KP} are divergent and the discussion in this paper is formal: We should investigate the energy density $\mathcal{E}(L)$ (the eigenvalue of the Hamiltonian density $\mathcal{H}(L)$) instead of the total energy $E(L)$  (the eigenvalue of the Hamiltonian ${H}(L)$).
In particular, we are to consider, instead of Eq.~\eqref{KP}, the commutator between $P_i$ and the boost operator denisity $\mathcal{K}_i (L)$, which may suffer from the singular Schwinger terms appearing  in the commutator among the components of stress tensor \cite{Boulware}.
Second, we have focused on the $n$-th mode $\phi^{(n)}$ in most of the present paper. We should sum up all of the modes and treat the resulting divergence using some regularization.
Detailed study of these two points  will be a subject of further research.
\section*{Acknowledgement}
We thank the anonymous referee for noticing the importance of the expectation values and commutators of stress tensor and the related references.
One of the authors (M. M.) thanks C. Hattori, M. Matsuda, and T. Matsuoka for discussion.



\end{document}